\documentclass[aip,amsmath,amssymb,jap,preprint,showpacs]{revtex4-1} 

\usepackage{graphicx}

\newcommand{\bj}{\mathbf{j}}
\newcommand{\bsig}{\mathbf{\boldsymbol\sigma}}
\newcommand{\sbkp}{\mathbf{k}_{\parallel}}
\newcommand{\bkp}{\mathbf{k}_{\parallel}}
\renewcommand{\Re}{\textrm{Re}}
\renewcommand{\Im}{\textrm{Im}}
\newcommand{\Tr}{\textrm{Tr}}
\newcommand{\tr}{\textrm{tr}}

\begin{document}
\title{New mechanism for generating spin transfer torque without charge current}
\author{G. Aut\`es$^{1,2}$}
\author{ J. Mathon$^{1}$, and A. Umerski$^{2}$}
\affiliation{$^{1}$Department of Mathematics, City University, London EC1V 0HB,
U.K.
\\ $^{2}$Department of Mathematics, Open University, Milton Keynes MK7 6AA,
U.K.}

\date{\today}

\begin{abstract}
A new physical mechanism for generating spin-transfer torque is proposed. It
is due to interference of bias driven nonequilibrium electrons incident on a
switching junction with the electrons reflected from an insulating 
barrier inserted in the junction after the switching magnet. It is shown using the
rigorous Keldysh formalism that this new out-of-plane torque $T_{\perp}$
is  proportional to an applied bias and is as large as the torque in a conventional 
junction generated by a strong charge current. However, the charge
current and the in-plane torque $T_{\parallel}$ are almost completely suppressed
by the insulating barrier. This new junction thus offers the highly applicable possibility of 
bias-induced switching of magnetization without charge current.
\end{abstract}


\maketitle
Slonczewski \cite{slon} proposed a new method of switching the magnetization
direction of a thin film by means of a spin-polarized current. The current is
spin-polarized by passing through a thick polarizing magnet
(PM),
whose magnetization is assumed to be pinned, subsequently passing through
a nonmagnetic metallic spacer layer of $N$ atomic planes and then through
a thin magnetic switching
layer (SM) into a nonmagnetic lead. We shall assume that the PM is semiinfinite and that its magnetization 
lies in the $xz$ plane
at an angle $\theta$ to the $z$ axis.
 The magnetization of the SM is assumed to be parallel to the $z$
axis.
The spin polarized current (spin current) is partly or fully
absorbed by the SM and the corresponding torque exerted on the SM can either
switch
its magnetization
completely or lead to steady-state precession of the
magnetization\cite{exp1,exp2}.
The current induced
precession of magnetization results in microwave generation. Both effects have
great potential for applications 
but the current density
required for magnetization switching in a conventional junction, shown
schematically in Fig.1a, is at present too large for commercial applications.

It is easy to see that there
is an upper limit on what can be achieved with conventional switching
junctions.
The maximum spin current is obtained when all carriers are 100\% spin
polarized, and
typical epitaxial junctions are already quite close to this theoretical limit.
One way to reduce the current flowing through the switching magnet is to use 
a three-terminal device\cite{Brataas}. However,
 a strong charge current still needs to be passed between the electrodes not
involved in switching. 
The quest for a system in which no strong charge current flows anywhere in the system
thus continues.

\begin{figure}[tp]
\begin{center}
\begin{tabular}{cc}
\includegraphics[width=4.2cm]{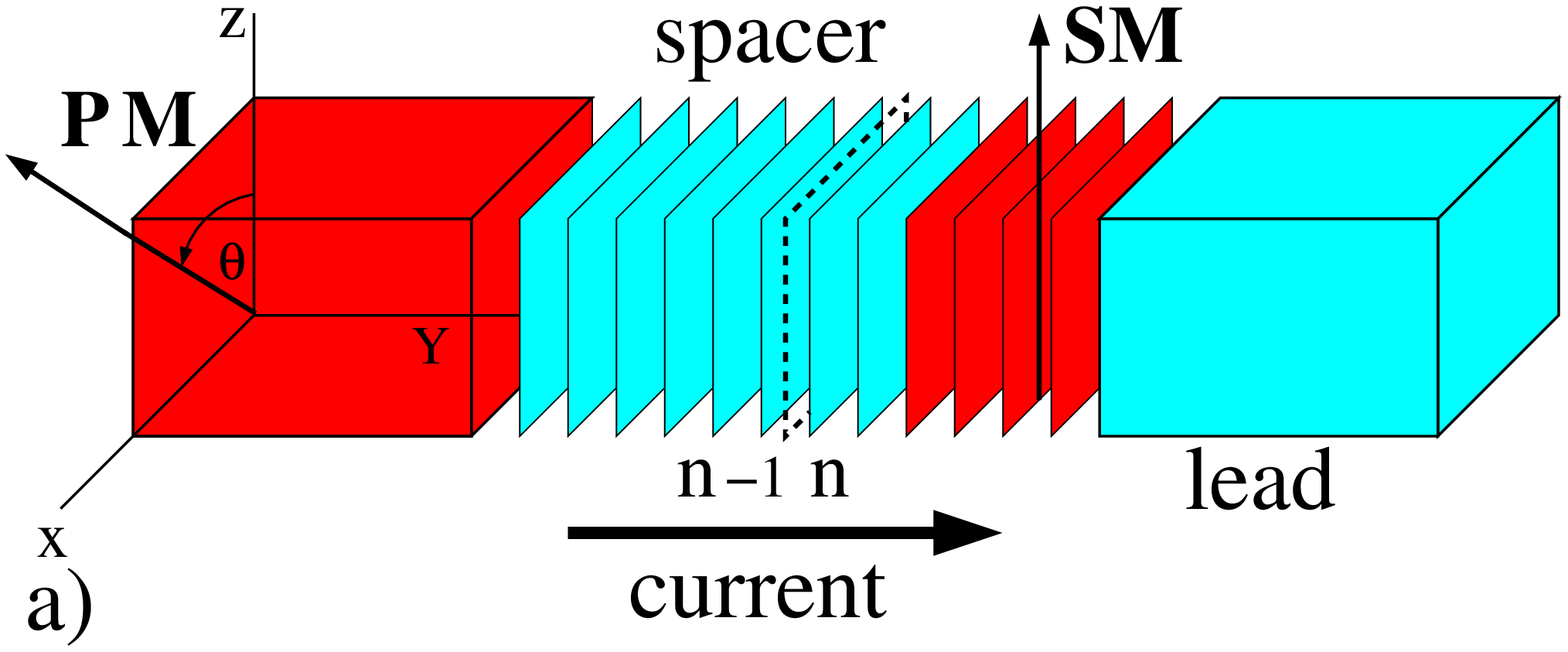}
&\includegraphics[width=4.2cm]{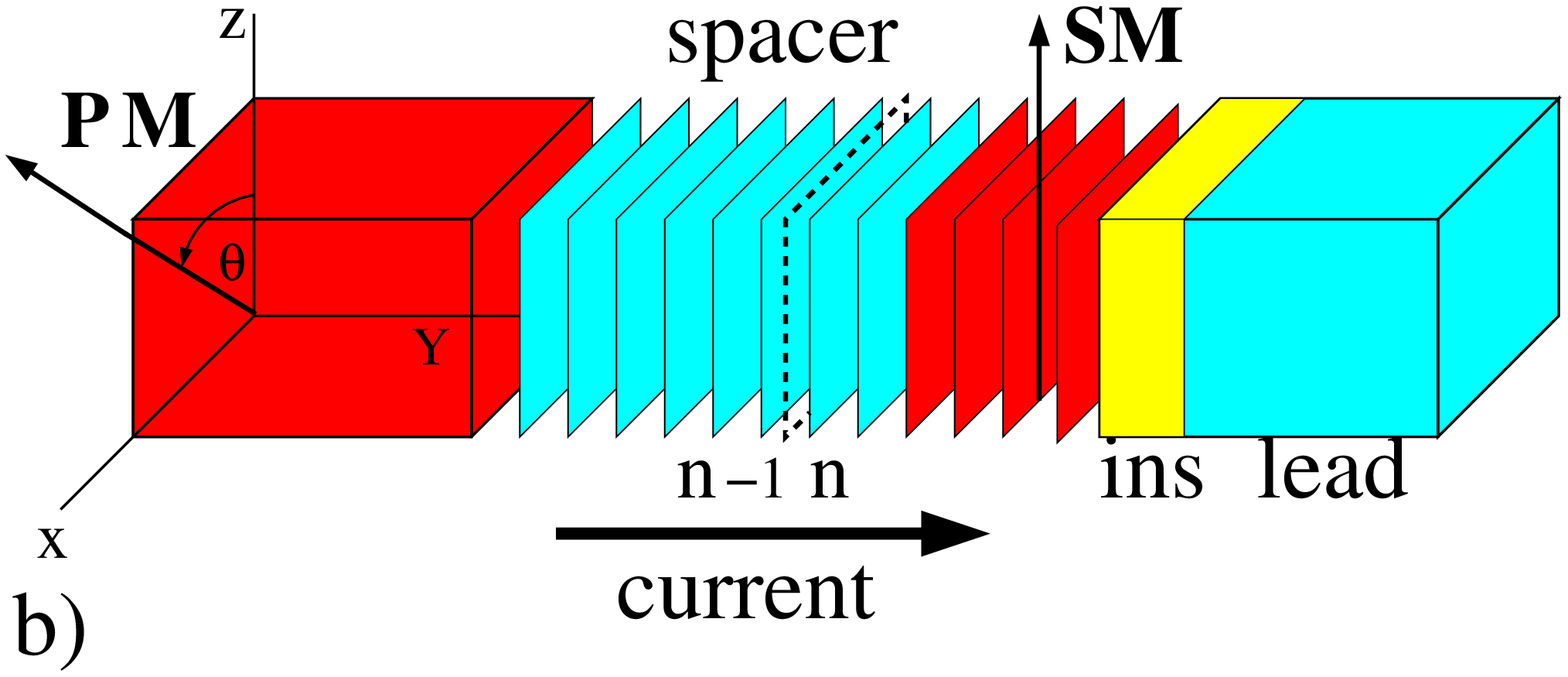}
\end{tabular}
\end{center}
\caption{\footnotesize (Color online) Conventional switching junction (a) and
the
junction with an insulating reflector (b).}
\label{fig1}
\end{figure}

We propose that a very large reduction of the switching current can be achieved
with a modified two-terminal junction shown in Fig1.b. The fundamental difference here is
that
a thin insulating layer is inserted between the switching magnet and the right
lead.
The charge current in such a junction is strongly reduced since it
has to pass through a tunneling barrier. However, we shall show that one of the
components of
the spin current in the nonmagnetic spacer layer is only weakly affected by the
barrier
and remains large even when the barrier is thick. One can, therefore, generate
a large
spin-transfer torque with a very weak charge current. 
To calculate it, we shall use a rigorous theory of the spin current\cite{edwards05}   based on the Keldysh
nonequilibrium formalism \cite{Keldysh} applied to a single-orbital tight-binding model
with nearest
neighbor hopping $t$ and atoms on a simple cubic lattice. Generalization to a
fully realistic
band structure is straightforward and is described in
\onlinecite{edwards05}.


The Keldysh formalism gives us a completely rigorous prescription of how to
calculate the steady-state spin and
charge current from the equilibrium retarded one-electron Green's functions
$g_{L}$ and
$g_{R}$ at the left and right surfaces of a junction cleaved between the planes $n-1$ and $n$.
It follows from Ref.\cite{edwards05} that the total spin current between atomic planes $n-1$, $n$  is the sum  of the equilibrium 
(zero bias) term $\bj^{0}_{n}$
and nonequilibrium (transport) term $\bj^{\tr}_{n}$
\begin{equation}
\bj _{n}^{0} = \frac{1}{4\pi}\sum_{\sbkp} \int d\omega \, \Re \Tr\{(B-A)\bsig
\}[f(\omega-\mu_{L})+f(\omega-\mu_{R})]
\label{eq1}
\end{equation}
\noindent
\begin{equation}
\begin{split}
\bj _{n}^{\tr} =& \frac{1}{2\pi}\sum_{\sbkp} \int d\omega \, \Re
\Tr\{[g_{L}tABg^{\dagger}_{R}t^{\dagger}-AB+ \\ &  \frac{1}{2}(A+B)]\bsig
\}[f(\omega-\mu_{L})-f(\omega-\mu_{R})].
\end{split}
\label{eq2}
\end{equation}
\noindent
Here, $A=[1-g_{R}t^{\dagger}g_{L}t]^{-1}$,
$B=[1-g_{R}^{\dagger}t^{\dagger}g_{L}^{\dagger}t]^{-1}$,
and $f(\omega-\mu)$ is the Fermi function with chemical potential $\mu$ and
$\mu_{L}-\mu_{R}=eV_{b}$.
The summation in Eqs.(1) and (2) is over the in-plane wave vector $\bkp$ and $\bsig$
is the Pauli matrix.
The charge current is calculated by replacing $\bsig$ with the unit matrix.
Since we only consider $\bj_n$ in the spacer where it is conserved, we drop the subscript $n$.

In zero bias only the equilibrium component of the spin current $j_{\perp}^{0}$ perpendicular
to the
plane determined by the PM and SM magnetizations ($xz$ plane) is nonzero. It
gives the
equilibrium interlayer exchange coupling \cite{edwards05}. It should be noted
that all occupied electron 
states contribute to the equilibrium coupling which is why Eq.(1) involves the
integral with respect
to energy. However, the equilibrium term (1) makes no contribution to the spin
current linear in the bias, i.e. to first order in $V_b$.
In the context of current-induced switching we can thus ignore this term and
focus on the transport contribution
given by Eq.(2). To the lowest order in the bias (linear response), the Fermi
functions in Eq.(2) are
expanded to first order in $V_{b}$. 
Hence the energy integral is avoided, being
equivalent to multiplying the integrand by $eV_{b}$ and evaluating it at the
common zero-bias chemical potential $\mu_{0}$. This shows explicitly  that
only states at the Fermi surface contribute, i.e., the term (2) is the nonequilibrium
transport contribution to the spin current.

It is now well known (see e.g. \cite{edwards05,butler}) that the transport spin
current in
the NM spacer of
a conventional switching junction (Fig.1a) has both in-plane $j_{\parallel}^{\tr}$
and out-of-plane $j_{\perp}^{\tr}$ components.
It has been argued (see e.g. Ref.\cite{butler}) that $j_{\perp}^{tr}$ linear in $V_{b}$
vanishes so that this term exhibits a quadratic dependence on the
applied bias. This is only true for a completely symmetric junction
but not true for asymmetric junctions as originally pointed out in \cite{edwards05} and later confirmed in \cite{tang}. Since the junction we propose
(see Fig. 1(b)) is inherently highly asymmetric, $j_{\perp}^{tr}$ linear
in $V_{b}$ is nonzero, and it is this term linear in the applied bias
which determines the transport out-of-plane torque.


Since the
magnetization of the PM is in the $xz$ plane the existence of the in-plane
spin current $j_{\parallel}^{\tr}$ is obvious but the origin of the out-of-plane
component $j_{\perp}^{\tr}$
is less clear. 
Electrons emerging from the PM magnet have spin polarized in the xz-plane and, therefore, 
$j_{\perp}^{\tr}$ can only arise in a FM/NM/FM junction as a result of 
reflections from FM/NM interfaces.

This observation has lead us to consider a modified junction shown
in Fig.1b in which an insulating layer (INS) is inserted between the SM and the right hand lead. 
When a bias is
applied across the junction, bias driven electrons at the Fermi level incident from the
left are strongly reflected at the SM/INS interface. The incident and reflected
electron waves interfere to form almost perfect standing waves. To a very good approximation, they are 
described by a real \cite{note} 
wavefunction $\Psi$ with components $\psi_{\uparrow}$ and $\psi_{\downarrow}$.
Since $j_{\parallel}^{\tr}\propto
\Im(\psi_{\uparrow}^{*}\psi_{\downarrow}'-\psi_{\uparrow}^{*'}\psi_{\downarrow})$ 
and $j_{\perp}^{\tr}
\propto \Re(\psi_{\uparrow}^{*}\psi_{\downarrow}'-\psi_{\uparrow}^{*'}\psi_{\downarrow})$, 
it is obvious that the corresponding
$j_{\parallel}^{\tr}$ vanishes identically for real $\Psi$. The same argument applies to the charge
current. On the other hand, $j_{\perp}^{\tr}$ is nonzero
for a standing wave.
It follows that the charge current and $j_{\parallel}^{\tr}$ are strongly
suppressed by the insulating layer but we expect that  $j_{\perp}^{\tr}$ remains
large and can even be enhanced by the insulating `reflector'. We emphasize
that, for this effect to occur, it is crucial that the `reflector' is placed
behind the switching magnet.
This is essential because incident and reflected electrons must travel across
the whole trilayer and feel spin-dependent potentials of both the PM and SM. 
We call the junction in Fig.1b a reflecting junction.

The total transport spin current can be again evaluated from Eq.(2) where the
surface Green's functions
$g_{L}$ and $g_{R}$ now include the effect of electron reflections at the
SM/INS interface. In Fig.2 we plot the spin currents
$j_{\parallel}^{\tr}$  and $j_{\perp}^{\tr}$, and the charge current $j_c$ as a function of the insulating barrier
thickness $N_{\textrm{INS}}$.
The angle between the magnetization of the PM and SM
layers is taken to be $\pi/2$ and the thickness of the SM is 5 atomic planes. 
We have used
the following values of tight-binding on-site potentials measured in units of $2t$: 
-2.3 and -2.8 for
the majority and minority spin in the PM and in the SM, -2.0 in the spacer and the lead, and -3.1 in the insulating barrier.
The thickness of the spacer is $N=12$ atomic planes.
Such a choice of parameters models a  Co/Cu/Co junction with a good matching of Co majority 
band with the Cu bands. For comparison, we  include in Fig.2 also the results for a 
conventional switching
junction corresponding to the insulating barrier thickness $N_{\textrm{INS}}$=0. 

\begin{figure}[tp]
\begin{center}
\includegraphics[width=6.0cm]{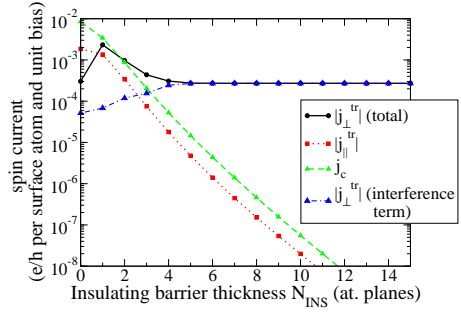}
\end{center}
\caption{\footnotesize In-plane ($j_{\parallel}^{\tr}$) spin current, out-of-plane
($j_{\perp}^{\tr}$) spin current, and charge current ($j_c$) in the spacer as a function of the
insulating barrier thickness. The magnetizations of the PM and the SM are perpendicular.
A conventional junction corresponds to $N_{\textrm{INS}}=0$.}
\label{fig2}
\end{figure}

It can be seen that for a conventional junction ($N_{\textrm{INS}}$=0), the in-plane and out-of-plane spin
currents are comparable in magnitude. However, the situation changes dramatically when an
`reflector' is inserted behind the switching magnet and the right lead.
The in-plane component  $j_{\parallel}^{\tr}$ and the charge current decrease exponentially
with the barrier thickness but the out-of-plane component $j_{\perp}^{\tr}$ saturates to a 
finite value which is quite close to the value of $j_{\perp}^{\tr}$ (and $j_{\parallel}^{\tr}$)
for a conventional junction.
To understand these results, it is important to note that
there are two different contributions to the out-of-plane spin current
$j_{\perp}^{\tr}$ in the NM spacer.
The first contribution is associated with the tunnelling charge current which carries 
with it an out-of-plane spin current component.
This is the usual out-of-plane component of the spin current 
which is observed in conventional switching junctions.
It is proportional to the charge current and thus decreases
exponentially with the barrier thickness.

The second (interference) contribution to  $j_{\perp}^{\tr}$ arises from interference between the
incoming and reflected electron waves. It is shown in Fig.2 as triangles. It can be seen that it is the only 
contribution that remains finite for thick insulating layer.
It arises because the bias driven electrons are almost totally reflected at the SM/INS interface and, therefore, 
almost perfect standing waves are formed in the NM spacer.
The origin of the out-of-plane spin current can then be explained using the following simple model 
of a standing wave
\begin{equation}
\Psi = \left(  \begin{array}{c}  A_{\uparrow} \cos(k_\perp y) \\
A_{\downarrow} \cos(k_\perp y + \phi)    \end{array}  \right),
\label{eq3}
\end{equation}
where the coefficients $A_{\uparrow}$ and  $A_{\downarrow}$ are real,
$k_\perp$ is the perpendicular wave vector in the NM spacer, and $y$ is the position in the spacer.
The phase shift $\phi$ between the majority- and minority-spin wave functions is a function of $k_\perp N a$, where $N$ is the spacer thickness and $a$ is the lattice constant.
The phase shift results in an
out-of-plane component of the spin current
\begin{equation}
j^{\tr}_{\perp}=k_\perp A_{\uparrow} A_{\downarrow}
\sin(\phi).
\label{eq4}
\end{equation}

For a given electron state with a parallel wave vector $\bkp$, the interference contribution
to the total out-of-plane spin current
oscillates around zero as the spacer thickness increases. The oscillation period is given by
$\pi/k_\perp(\sbkp)$.
The total $j^{\tr}_{\perp}$  is obtained by summing over all
$\sbkp$ states in the 2D Brillouin zone (see Eq.\ref{eq2}). States with
different $\sbkp$ have different oscillation periods and, therefore, interfere destructively. 
It follows \cite{Itoh} that the oscillation amplitude 
of the integrated interference component of $j^{\tr}_{\perp}$  decreases
with increasing spacer thickness.
The total out-of-plane spin current is thus expected
to oscillate with a decaying amplitude
about a small constant background determined by the tunneling component. 
Since the magnitude of the spin current in the spacer of a reflecting junction decreases with
the spacer thickness we need to establish that, for a realistic bias and realistic spacer thickness, 
the resultant torque
on the switching magnet is at least as large as in a coventional junction and also that the transport
torque $T^{\tr}_{\perp}$ is stronger than the equilibrium interlayer coupling torque $T^{0}_{\perp}$.
The torque exerted on the switching magnet is the difference between the spin currents in
the spacer and right lead. To evaluate the torque, we note that the transport 
spin current  in the right lead has only the tunneling component of $j^{\tr}_{\perp}$ 
which is negligible compared with
the interference component of $j^{\tr}_{\perp}$ in the spacer. The equilibrium spin current 
$j^{0}_{\perp}$ in the lead is strictly zero. It follows that both torques
$T^{\tr}_{\perp}$ and $T^{0}_{\perp}$ are given by the corresponding spin currents in the spacer.
The fact that $T^{\tr}_{\perp}$ in a reflecting junction with a spacer thickness of the order of 10 atomic
planes can be as large as the torque in a conventional junction is already evident from Fig.2.
We, therefore, only need to compare the transport torque  $T^{\tr}_{\perp}$ with the equilibrium 
coupling torque $T^{0}_{\perp}$. Since the transport torque is proportional to the bias $V_{b}$,
it is necessary to choose for this comparison a value of $V_{b}$ small enough for the linear-response 
approximation
adopted here to be valid. A value $eV_{b}=w/600$, where $w$ is the band width, satisfies this requirement
since the voltage drop across the barrier is negligible compared with the barrier height
($\approx w/2$ in Fig.2). We note that, to the lowest order (linear) in $V_{b}$, the equilibrium 
coupling torque is independent of the bias. Using $eV_{b}=w/600$, which corresponds to a bias $V_{b}=0.01V$
for  $w=6eV$, we compare in Fig.3 the transport and equilibrium coupling torques
assuming that the angle between the PM and SM magnetizations is $\pi /2$. The tight binding
parameters are the same as in Fig.2.

\begin{figure}[tp]
\begin{center}
\includegraphics[width=6.0cm]{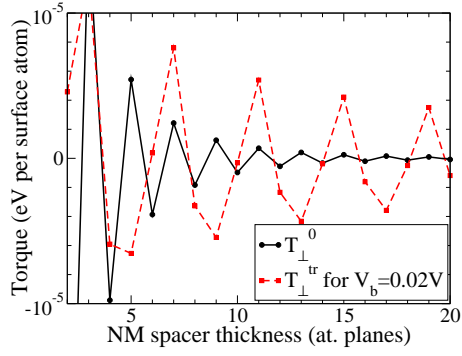}
\end{center}
\caption{\footnotesize interlayer coupling torque ($T^{0}_{\perp}$) and transport torque ($T^{\tr}_\perp$) as a function of the NM spacer thickness, when the magnetization of the PM and SM are orhtogonal and the applied bias is $V_b=0.01V$. The units are assuming a band width $w$ of $6eV$ (i.e. $t=0.5eV$).}
\label{fig3}
\end{figure}
\noindent

Both torques oscillate with decreasing amplitude as the thickness of the spacer
increases.
However, the amplitudes, periods, and decay rates of the equilibrium and transport torque oscillations 
are quite different, which clearly demonstrates their fundamentally different origins.
We first note that, even for the very low bias of $V_{b}=0.01V$, the transport torque is much 
stronger than the coupling torque. There is, therefore, no problem in overcoming the static coupling
term by the bias-dependent transport term. Moreover, since the two torques oscillate with different
periods, one can always select a spacer thickness where the static coupling is close to zero and thus 
eliminate this term altogether.

We now briefly discuss the oscillation periods and decay rates of $T^{0}_{\perp}$ and $T^{\tr}_{\perp}$.
It is well known \cite{coupling} that the static torque  $T^{0}_{\perp}$  decays as
$1/N^2$, where $N$ is the thickness of the spacer. The corresponding oscillation period is given
by the spacer Fermi surface (FS) spanning vector \cite{coupling} (2 atomic planes in our case). The
periods obtained from the extrema of the spacer FS are the only periods that can occur for the
equilibrium coupling torque \cite{coupling}. However, the transport torque can also oscillate with 
additional periods
arising from sharp cutoffs of the sum over $\bkp$ in Eq.(2) (the cutoff periods are removed from
the equilibrium coupling term by the energy integral in Eq.(1) \cite{Itoh}). 
The origin of the cutoff periods
was discussed by Mathon et al.  \cite{Itoh} in the case of charge current oscillations, and
the same arguments apply here. 
Finally, the decay of the transport torque oscillations with spacer thickness should be 
slower than the $1/N^2$ decay rate of the static coupling. This is because
the additional destructives interference that arises from the energy integration in the
static coupling term (Eq.\ref{eq1}) is not
present in the transport term (eq.\ref{eq2}).
In the case shown in Fig.3, the oscillation period of the transport torque $T^{\tr}_{\perp}$ 
is clearly dominated by a cutoff period which  is $\approx 4$ atomic
planes for the potentials we have chosen. 
The decay rate of $T^{\tr}_{\perp}$ is slower than that of the coupling torque (see e.g. \cite{Itoh}). 

Finally, we point out that although our results are for a switching (SM) thickness of 5 atomic planes, qualitatively similar results are obtained for other SM thicknesses.
Varying the thickness of the SM has only a small effect on the transport torque $T^{\tr}_{\perp}$, i.e. it oscillates  with a small amplitude around a finite constant background as the SM thickness increase. This is because most of the interference responsible for  $T^{\tr}_{\perp}$ occurs in the spacer.

The reflecting junction we propose offers huge potential advantages over the conventional junction.
Firstly, a strong out-of-plane spin-transfer torque can be generated by an applied bias without
the accompanying charge current. The bias strength is not limited to the linear-response regime
considered here. Generalization to a strong bias simply requires energy integration in Eq.(2) between
$\mu _{L}$ and $\mu _{R}$. The applied bias is then limited only by the barrier height. The second
advantage of the reflecting junction is that the magnitude and sign of the ratio 
$T^{\tr}_{\perp}/T^{\tr}_{\parallel}$ can be tuned by the height/width of the reflecting barrier
and by the spacer thickness. This is important since the ratio $T^{\tr}_{\perp}/T^{\tr}_{\parallel}$
controls switching scenarios \cite{E+M}. For example, with the appropriate sign of this ratio, 
microwave generation can be achieved without an applied magnetic field \cite{E+Mreview}.

A bias controled switching was proposed earlier in \cite{suzuki}. However,
the physical mechanism behind this idea is completely different. It is based on a bias induced
modification of the equilibrium interlayer coupling and ignores completely the transport term considered
here. However, as already discussed, the modification of the equilibrium coupling by a bias is
a higher order effect which vanishes to the first order in the bias.

Since the out-of-plane torque $T^{\tr}_{\perp}$ arises from interference between incident and
reflected electron waves one needs good interfaces to observe and exploit it. However, the quality
of the interfaces need not be any better than that required for observation of the usual interlayer
exchange coupling, which is also an interference effect. 
In addition, the quality of the SM/INS
interface may also be important.
 However since the main role of the insulator is to suppress the charge current, the quality of this interface may not be so crucial.
Furthermore, it is known from experiments on tunneling junctions with MgO
barrier that the Fe/MgO interface can be grown almost perfectly epitaxial, and we suggest that
this combination would be an ideal choice for the reflecting junction.
Finally, we would like to mention that an insulating barrier could be replaced by a doped
semiconductor layer such as InAs which forms an ohmic contact with SM (e.g. Fe). This
might allow a finer tuning of the ratio $T^{\tr}_{\perp}/T^{\tr}_{\parallel}$ since 
the spin current $T^{\tr}_{\parallel}$ that can flow through the junction could be
controlled by doping (size of the semiconductor FS). 

\begin{acknowledgments}
We are grateful to the UK Engineering and Physical Sciences Research Council
for financial support.
\end{acknowledgments}



\begin{thebibliography}{99}
\bibitem{note} Note that $\Psi$ is never strictly real. It would be real only
for an infinitely high insulating barrier, in which case the junction would cease to be a
nonequilibrium system.
\end{thebibliography}
\end{document}